# Evidence for a Galactic Origin of Very Short Gamma Ray Bursts and Primordial Black Hole Sources

D.B. Cline, C. Matthey and S. Otwinowski
University of California Los Angeles
Department of Physics and Astronomy
Box 951447, Los Angeles, California 90095-1547 USA

## Abstract

We systematically study the shortest time duration gamma ray bursts and find unique features that are best interpreted as sources of a galactic origin. There is a significant angular asymmetry and the $V/V_{max}$ distribution provides evidence for a homogenous or Euclidean source distribution. We review the arguments that primordial black hole evaporation can give such GRBs. The rate of events is consistent with a PBH origin if we assume on enhanced local density, as are the other distributions. We suggest further tests of this hypothesis.

## 1. Introduction

The earliest observations of very short (100 ms or less duration) gamma ray bursts was made in 1979 – 1991 [1,2]. It was even remarked that these could be a separate class of GRBs from the better known events with much longer time duration [1]. In the 1980s the interplanetary network used to locate the position of the GRB origin was superior to that of the latter times. For the very short bursts careful observation never revealed a source at the location [1]. For the longer bursts we now know that this was due to the fact that nearly all of those sources are cosmological with respect to galactic sources [3,4,5]. However, as we will show in this paper the very short GRBs could well arrive from galactic sources so the lack of any detected origin to the events is possibly more significant. In the 1990s the BATSE detector on the Compton Observatory recorded many more very short GRB's, which are the subject of this paper [6].

In 1971 the concept of the primordial black holes produced in the early universe was introduced [7]. Subsequently the possibility of Hawking Radiation was derived [8]. However there has never been any firm evidence for the existence of Primordial Black Holes. Application of Hawking Radiation formula indicates that all PBH with mass of less than $5 \times 10^{14}$ grams would have already evaporated in the universe [9]. Furthermore the photonic debris from the evaporation would give a diffuse $\gamma$ ray background in the universe [9][10]. This diffuse background then is used to set limits on the total number of the PBH in the universe. The number of PBH in a given galaxy or even the local region near the earth cannot be calculated directly with current knowledge. It is possible that the PBH forms like cold dark matter is sometimes suggested to do or could be concentrated around old stars etc. [11].

Thus the search for local sources of PBH must start with an unbiased viewpoint about the local density given the constraints of the overall number of PBH in the universe. If we assume that PBH's cluster in the galaxy the way baryons do then the usually quoted average upper limit of PBH per cubic parsec is [10,12]

$$N_{PBH}^{G.Halo} \leq \text{PBH}/(\text{pc})^3 .$$

However this density could be more or less in different parts of the Galaxy [11,12].

In this paper we first report a final analysis of some properties of the very short gamma ray bursts and show that they could well comes from a Galactic source – we then make the hypothesis that the origin of this GRB's could be nearby Primordial Black Holes [13][14]. We have previously reported on a partial sample of these data [13].

Finally we propose future tests that could determine if these sources are Primordial Black Holes.

## 2. Gamma ray bursts from PBH evaporation

As Hawking Radiation proceeds the PBH loses mass, the temperature increases. The temperature of the PBH is $T = m_p^2 / 8\pi m$, where $m_p$ is the Planck mass and M the PBH mass in grams.

$$T = \frac{10^{10} \text{gm}}{M} \text{ TeV} \quad [9,10].$$

There are two regions where gamma ray bursts could appear:

(1) As the temperature reaches this QCD – Quark - Gluon Phase transition
    (T ~ 160 GeV) a fireball formation could occur [13,14][15].

(2) As the temperature reaches the 10 GeV to TeV range, a plasma will form near the PBH. The evaporation wind can form a magnetic field and MHD instability can lead to burst of gamma rays [16].

The mass/energy loss of PBH is given by

$$\frac{dM}{dt} = \frac{\alpha(M) M_p^2}{M^2}$$

where $\alpha(M)$ is the function reflecting the species of particles created by the black hole gravitational field and M is mass of the PBH.

$\alpha(M)$ grows with temperature and will have a rapid change at a first order QCD phase transition (~160 MeV). [13,14,15]. If we integrate over time we find that the lifetime of all black holes with mass less than 5 x $10^{14}$ gram will have a lifetime less than the Universe so they would have evaporated by now. PBH with mass 5 x $10^{14}$ gram would be evaporating at this time in the universe. The horizon mass of the universe that had this value mass is $10^{-23}$ sec after the Hot Big Bang [10,12]. So if these PBH were made at the horizon time they could be among the oldest objects in the universe.

Returning to the GRB that could be emitted by the evaporating PBH we note that in neither case (1) or (2) are there fully developed calculations of the expected time duration, energy spectrum or photon luminosity of the GRB. However there are reasonable estimates that low energy photons (MeV) with luminosity of $10^{34}$ergs/cm²/sec (case 1) to $10^{30}$ergs/cm²/sec (case 2) [16] would be formed. The small size of the PBH plasma would suggest that the GRBs have shorter time duration. Experimental study of PBH evaporation will likely be required to fully understand this process. The detection rate for PBH in either (case 1) or (case 2) will depends on the local density of PBH. Since there are no direct constraints on this density, and clustering would occur as in CDM Models again, this is subject to experimental study [11,12].

Given the uncertainty of PBH evaporation, clustering and formation in the early universe, it is essential to find signals for PBH evaporation in the experimental data. There are three general expectations:

(1) the PBH should be galactic sources;
(2) the PBH should strongly cluster to provide significant local density;
(3) the PBH should give unique nearly identical GRBs event due to simplicity of the conditions and fact that all PBH evaporating today start at the same temperature.

## 3. Estimated Detection Rate for GRBs from Primordial Black Holes Evaporation

The two-temperature region where GRBs may be formed from PBH evaporation can have very different luminosity. Estimates for the number of events that might be detected by BATSE have been given previously for the fireball model with temperature near the Q G phase transition. In Fig. 1 we show a flow diagram that relates the various constants on the rate from different measurements [17]. The estimated rate of events is up to 10 per year in BATSE. For the $2^{nd}$ model discussed before the authors also estimate [5-10] per year [16]. These estimates are not likely to be accurate but do indicate that in these models a few events per year may be detected by BATSE. There are excellent reviews of the status of primordial black holes in Ref. 18 and 19 and in reference [12].

## 4. Detailed Study of Very Short of Duration Gamma Ray Bursts from BATSE

The time distribution T90 for all GRBs from the BATSE detector up to May 26, 2000 shows two clear peaks and a small excess below 100 ms duration. We divide the GRBs into three classes in time duration: L ($\tau > 1$ s) long; M ($1\sigma > \tau > 0.1$ s) medium; and S ($\tau < 100$ ms) very short [20]. The duration time of T90 is used for all of this analysis. In this paper, we confine the discussion to the M and S classes of GRBs. Since these events are adjacent in time, it is important to contrast the behavior.

We assume that the S GRBs constitute a separate class of GRBs and fit the time distribution with a three-population model. The fit is excellent but does not in itself give significant evidence for a three population model. We now turn to the angular distributions of the S and M GRBs. In Fig. 2a, we show this distribution for the very short bursts (46 events in total). We can see directly that this is not an isotropic distribution. To a certain the significance of the anisotropy, we break up the Galactic map into eight equal probability regions. In Fig. 2a, we show the distribution of events in the eight bins; clearly one bin has a large excess.

To contrast the distribution of the S GRBs and to test for possible errors in the analysis, we plot the same distributions for the M sample in Fig2b. As can be seen from Fig.2b, this distribution is consistent with isotropy. Fig.2b shows the same analysis as Fig.2a, indicating that there are no bins with a statistically significant deviation from the hypothesis of an isotropic distribution. Of the 46 very short S GRBs, there are 20 in the excess region. We have studied events from S and M samples and can find no real differences between the properties of the excess events and those outside of the excess region.

We can make a preliminary conclusion based on Fig2a, and $< V/V_{max} > = 0.76 \pm 0.14$ that the S events are likely from Galactic, or possibly more local (solar neighborhood), sources. The $V/V_{max}$ parameter is related to the volume of space that the source resides divided by the maximum volume for such a source [6] and is related to the geometry of the sources: $V/V_{max} = \frac{1}{2}$ for a Euclidean geometry. This is the first convincing evidence of some GRBs that are probably at non-cosmological distances.

To further contrast the GRBs in S and M regions, we have calculated $V/V_{max}$ for each event using $C_p$ values (the event counts) from the BATSE data. [6] In Fig.3, we show the distribution of $V/V_{max}$ for S events. As we have previously noted ([13]), this distribution is totally consistent with $<V/V_{max}>) = 0.5$ for a local distribution. In contrast, the same distribution for M events show in Fig.2b indicate a $<V/V_{max}>$ much less than 0.5 consistent with the same mean values for the L (long) events, which is now widely interpreted as being due to the cosmological sources for those GRBs. It is probable that the M events are also from cosmological sources; however the S events appear to came from local sources. We note that the short bursts are strongly consistent with a Cp spectrum, indicating a Euclidean source distribution, as was shown previously by ([13]). In the medium (from 100 ms to 1 s) time duration, the ln N - ln S distribution seems to be non-Euclidean ([6]). The $<V/V_{max}>$ for the S, M, and L class of events is, respectively, $0.76 \pm 0.14$, $0.37 \pm 0.03$, three standard deviations different.

To determinate the statistical probability for such deviation, we calculate the Poisson probability for eight bins with a total of 46 events (see Fig.2a). The probability of observing 20 events in a single bin is $1.6 \times 10^{-5}$. We consider this as a very significant deviation from the isotropic distribution. We note that the M event distribution (Fig. 2b) is fully consistent with isotropy of the sources on contrast.

We perform also the likelihood analysis testing two hypothesis:

(h1) Poisson distribution with $\lambda = 5.75 = 46/8$

In this case the logarithmic probability ln(p) calculated for the experimental sample is ~ -3.13, while the average ln(p) estimated from $10^5$ randomly generated samples using the Poisson distribution with the same $\lambda$ and total number of events is ~ -17.76, with the standard deviation SD ~ 1.74. We conclude that the observed value is about 7.8 SD below the Poisson average. This corresponds to the ~ $1.4 \times 10^{-6}$ chance to observe such a configuration and discards (h1).

(h2) Poisson distribution with extra source in "anomalous" angular bin.

Testing (h2), we first estimate the hypothesis that the underlying distribution for 7 angular bins except the anomalous one is indeed Poisson with $\lambda = 3.714 = 26/7$. The ln(p) of the experimental sample is in this case ~ -15.29, which is less then one SD (~ 1.532) from the average for the random Poisson samples (~ -13.905). That confirms the Poisson hypothesis. One can get a rough estimate of the parameters characterizing (h2): $\lambda$ ~3.7, X~ 20 –3.7 = ~ 16. The direct minimization of the likehood for (h2) gives $\lambda$ ~ 3.78±0.7, X~ 15.7±1.5 in a good agreement with the previous estimate.

The likely minimum value is ~ -15.5 (which should be compared with 15.29).

We consider that this results are strong evidence for acceptance of (h2) and conclude that S GRBs are distributed isotropically with mean ~ 3.7, but there is an extra source which yields ~16 events in the anomalous angular region.

Independent of a direct association of the GRB events with specific galactic sources, it would be even harder to explain the distributions in Fig. 2a as being due to extragalactic or even cosmological sources. The value of the $<V/V_{max}>$ and the location asymmetry would seem to strongly support a Galactic origin of the sources for the S GRBs. It is also clear that this distinction would equally argue for a separate class of S GRBs from the M or L classes.

We believe a future study of the S GRB population so a possible shorter time distribution will be fruitful. We may only have detected a fraction of the short bursts ([21]). The bulk of the very short bursts identified here all have time duration at or below the BATSE 64-ms integration time [21]. We therefore believe that the BATSE trigger is likely an inefficient method of identifying such events; we also believe that many weak bursts may have been missed [21]. There could be as many missed GRBs of 1-ms duration as the number that have been detected at 100 sec. according to Reference 21.

## 5. Properties of Sources and the Data

If the very short GRBs are from galactic sources as we have argued above, there are some striking features of these events:

1) There is no excess from either galactic center or galactic plane in Figure 2a. Nearly all stellar process or processes involving collapsed stellar objects should have a strong concentration in the galactic center or plane.

2) The source appears to be distributed more as the halo or dark matter distribution except for the excess noted above.

3) There are some expectations that dark matter sources are clumped in the halo [22]. The density in the clump can exceed by a factor of 100 or more the average density [22]. One would expect some asymmetry in the angular distribution due to the unlikelihood of the solar system being in the exact center of any given cluster.

4) We note that there are no clear time or angular coincidences in the angular distribution or data, suggesting that the sources do not repeat over the 9-year operation of BATSE.

## 6. Future Tests for the PBH Hypothesis and Summary

Possible source contributors to the angular asymmetry of astrophysical sources:

1. It would seen unlikely for any extragalactic source to give such an asymmetric distribution.
2. We have studied various objects such as nearby stars and can find no angular excess in the same regions as Fig.2a. Future tests should confirm this search.
3. If neutron stars were producing very short duration GRB we would expect:

   (a) An excess at the galactic center
   (b) An excess at the galactic plane

   Neither of this excess appears in Fig.2a.
4. We conclude the excess likely arises from some others source such as Primordial Black Holes and may cluster in clumps like CDM.

The flux and detection of antiprotons from PBH evaporation and of possible higher energy particles is subject to some uncertainty due to the plasma that forms around the PBH. For a recent summary and past references see Ref. 23. Other possible tests of the PBH hypotheses are found in Ref. 11 and 16.

It has been pointed out that PBH binary systems could give GRBs in Ref. 24. These events could have the properties of the events discussed here. High frequency gravitational wave bursts would also be expected.

Future observations would test these hypotheses and search for a visible signal from the source that would seem to be unlikely if a PBH is the source.

## 7. Summary

We have presented evidence that very short gamma ray bursts constitute a separate class of GRBs due to:

(1) the angular asymmetry
(2) the $V/V_{max}$ distribution

and differs from all other classes of GRBs.

We have shown that the evaporation of a local population of primordial black holes could provide a source of these GRBs. We have suggested future tests of this hypothesis.

We thank C. Meegan for help with the BATSE data.

## References

1. C. Barat et al, 1984, ApJ **285**, 791.
2. P. Bhat et al., 1992, Nature, **389**, 217.
3. E. Costa et al., 1997, Nature, **387**, 783.
4. Van Paradijs et al., 1997, Nature, **386**, 686.
5. M. Metzger et al, 1997, Nature, **387**, 878.
6. A. Fishman et al., 1994, ApJ **92**, 229.
7. B. Zeldovich and I.D. Novikov, 1966, Sn Astro AJ **10**, 602.
8. S. Hawking, 1974, Nature **30**, 248.
9. D.N. Page and S.W. Hawking, 1976, ApJ, **206**
10. B. Carr and S.W. Hawking, 1974, MNRAS **168**, 399.
11. E.V. Derishov and A.A. Belyanin, 1999, Astron. Astrophys., **343**, 1.
12. See Phys. Reports **307**, 1998 (D. Cline editor) for a review of many of the issues about PBH.
13. D.B. Cline, C. Matthey and S. Otwinowski, 1999, ApJ **527**, 827-834.
14. D.B. Cline and W. Hong 1992, ApJ **401**, L57.
15. D.B. Cline, Nuclear Phys. A, 1996, **610**, 500.
16. A.A. Belyanin, V. Kocharovsky and VI. V. Kocharowsky, 1996, MNRAS **283**, 626.
17. D.B. Cline, 1998, ApJ **501**, L1.
18. J.H. MacGibbon and B.J. Carr, 1991, ApJ **371**, 447.
19. F. Halzen et al., 1991, Nature, **353**, 807.
20. D.B. Cline, D.A. Sanders, and W.P. Hong, ApJ **486**, 169 (1997).
21. R.J. Nemiroff, et al., ApJ. **494**, L13 (1998).
22. B. Moore et al., 1999, ApJ **524**, 19.
23. J. Cline et al., 1998, Hep-ph 99810439.
24. J.N. Abdurashitov, V.E. Yants and C.V. Parfenov, Astro-ph 9911093.

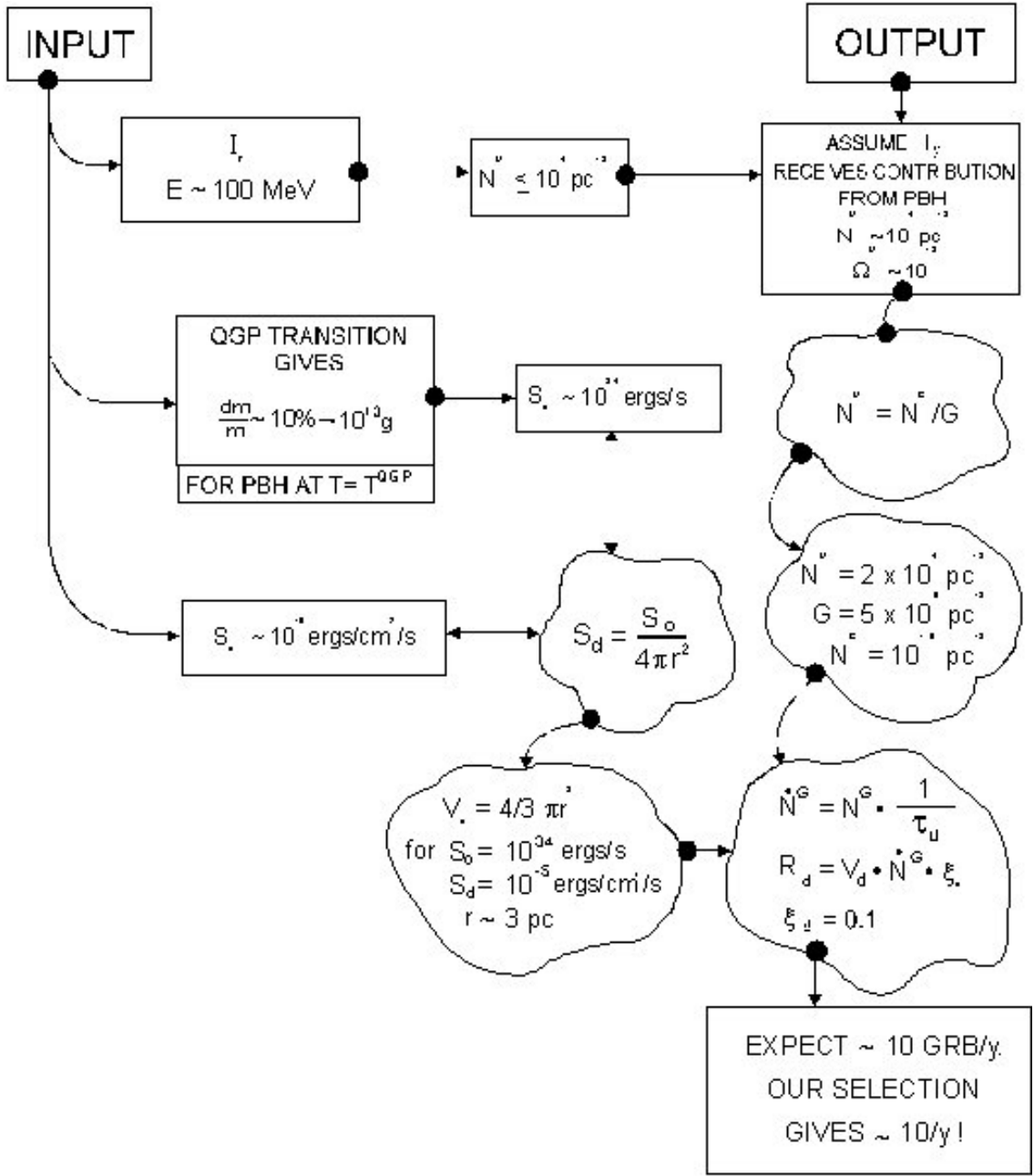

**Figure 1.** Schematic of the connection between GRB events and the diffuse γ background. L_γ where $N^U$ is density of PBH in the universe, $\Omega^U$ is ratio of density of the universe, $T^{QGP}$ is QGP transition temperature, $S_o$ is GRB luminosity from PBH, $N^0$ is the number of PBHs in the galaxy, G is galaxy clumping factor, $S_d$ is fluence sensitivity of GRB detectors, r is the distance to the source, $N^G$ is the rate of PBH decay in the galaxy, $\tau_U$ is the age of the universe, $R_d$ is the number of GRBs detected per year from PBHs, $\xi_d$ is GRB detector efficiency (including the fraction of energy detected).

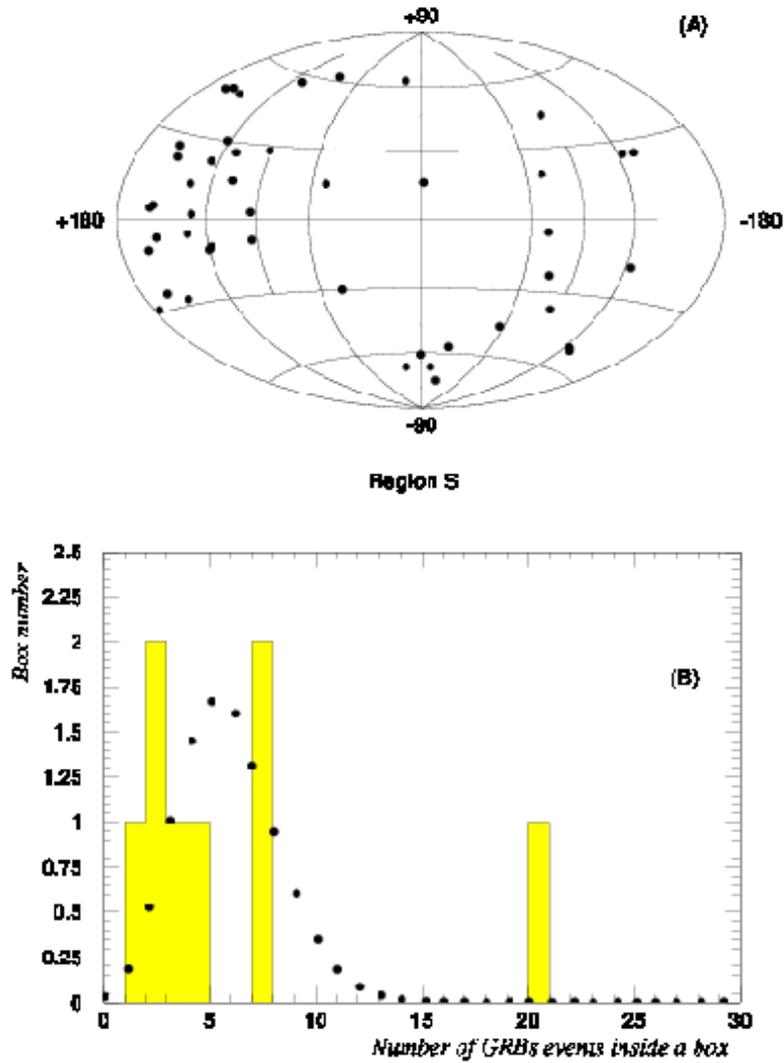

**Fig. 2a.** The Galactic coordinate angular distribution of the GRB events (upper graph) with very short time duration (S). We define eight regions that correspond to equal angular surface. The distribution of the number of GRB events (lower graph) in each of the eight regions. Points correspond to prediction of the Poisson distribution for 42 events divided into eight equiprobable groups. Both distributions are normalized to the same surface.

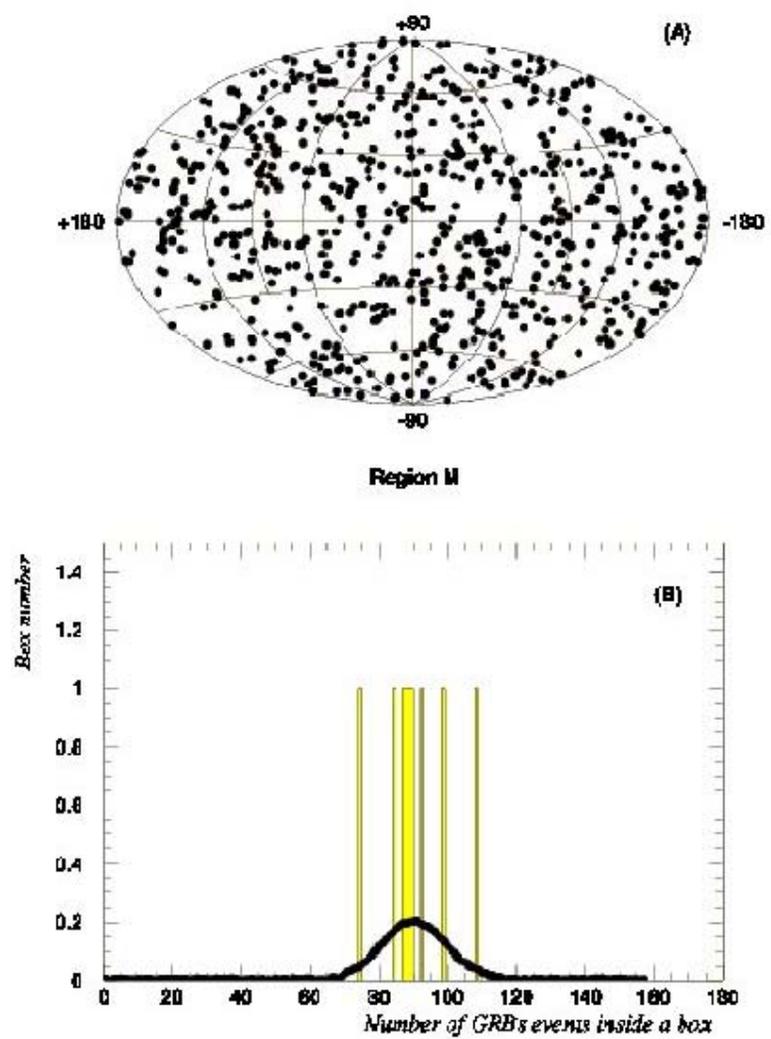

**Fig. 2b.** The Galactic coordinate angular distribution of the GRB events (upper graph) with short time duration (M). Eight regions correspond to equal angular surface (like Figure 2A). The distribution of the number of GRB events (lower graph) in each of the eight regions. Points correspond to prediction of the Poisson (Gauss) distribution for 269 events divided into eight equiprobable groups. Both distributions are normalized to the same surface.

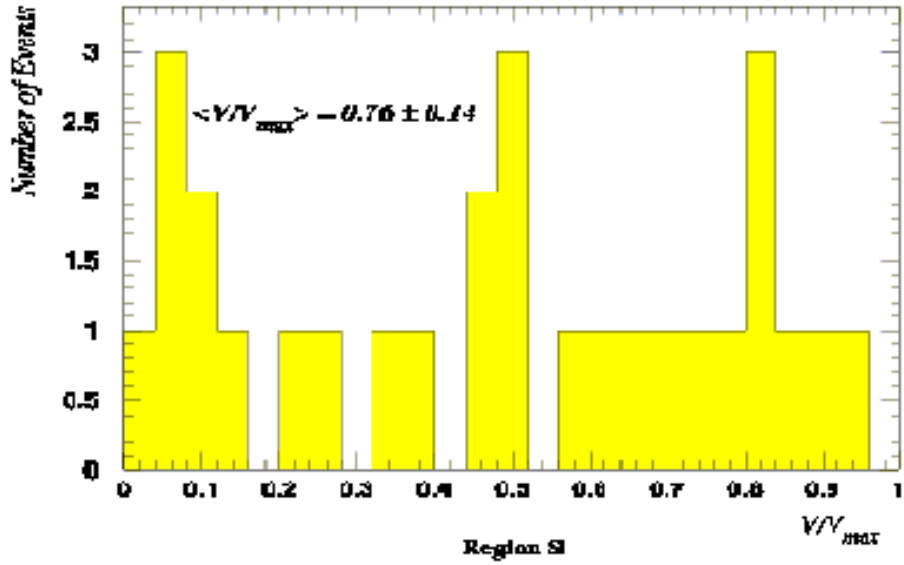
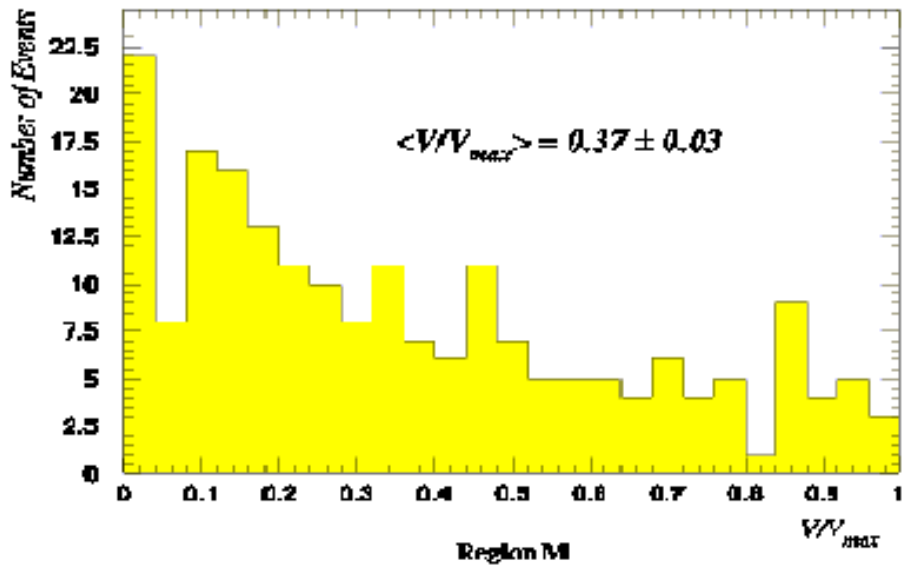

**Fig. 3.** $V/V_{max}$ distribution for (a) very short time duration, S; and (b) short time duration, M.